\documentclass[prb,aps,floats,superscriptaddress,floatfix,twocolumn,longbibliography]{revtex4-2}
\usepackage{gensymb}
\usepackage{graphicx}
\usepackage{amsmath, amsthm, amssymb, amsfonts}
\usepackage{hyperref}
\usepackage{array}
\usepackage{booktabs}

\begin{document}

\title{Coexistence of Griffiths Phase and Superparamagnetic-like
Behavior in { La-doped} Antiferromagnetic CaMnO$_3$}
 
 \author{Koustav Pal}
 \email{koustav.pal97@gmail.com, koustavpal001@scut.edu.cn}
 \affiliation{Spin-X Institute, School of Physics and Optoelectronics, State Key Laboratory of Luminescent Materials and Devices, Guangdong-Hong Kong-Macao Joint Laboratory of Optoelectronic and Magnetic Functional Materials, South China University of Technology, Guangzhou 511442, China}
 
\author{Shuvendu Ghosh}
\email{shuvendug@adityauniversity.in}
\affiliation{Department of Physics, Aditya University, Surampalem, 533437, Andhra Pradesh, India}

\author{N.N.S.V. Rama Raju}
\affiliation{Department of ECE, N S Raju Institute of Technology (NSRIT) Sontyam, Visakhapatnam, 531173, Andhra Pradesh, India}

\author{Chitrada Prasad}
\affiliation{Department of Physics, Aditya University, Surampalem, 533437, Andhra Pradesh, India}

\author{Kalipada Das}
\affiliation{Department of physics,Seth Anandram Jaipuria College, 10-Raja Nabakrishna Street, Kolkata-700005}	

\author{I. Das}
\affiliation{Saha Institute of Nuclear Physics, A CI of Homi Bhabha National Institute, 700064, Kolkata , India}

\date{\today}

\begin{abstract}
The coexistence of a Griffiths phase and superparamagnetic-like short range interaction in a single-phase homogeneous manganite is of particular interest because it reveals multiple levels of magnetic clustering generated by disorder and competing exchange  interactions. In bulk La$_{0.19}$Ca$_{0.81}$MnO$_3$, this unusual 
magnetic state develops in the antiferromagnetic regime, where A-site 
disorder strongly modifies the local magnetic landscape. Rietveld analysis 
confirms a single-phase orthorhombic structure, while magnetic measurements 
establish a Griffiths phase below $T_{\mathrm{G}} = 280$~K through the 
deviation from Curie--Weiss behavior.  Superparamagnetic- like behavior persisting 
from 280 to 370~K further indicates that isolated magnetic clusters remain 
stable well into the nominally paramagnetic region. The presence of both 
phenomena highlights the crucial role of disorder-induced exchange 
randomness in stabilizing complex magnetic states in correlated perovskite 
oxides.
\end{abstract}

\maketitle

\section{Introduction}

A Griffiths phase (GP) is defined as a short-range ferromagnetic (FM) ordered state that typically occurs in magnetic systems exhibiting quenched disorder, above the temperature at which long-range magnetic ordering sets in~\cite{ref1,ref2,ref3,a1}. The appearance of this phase, along with its characteristic singular behavior, is generally attributed to the complex interplay between magnetic energy barriers and thermal fluctuations present in the system~\cite{Griffiths1969,ref5,a2}. Although the exact mechanism behind the formation of the Griffiths phase in different magnetic materials is not yet fully understood, it is commonly linked to factors such as phase separation, competing magnetic interactions, the coexistence of clusters of varying sizes, and the presence of disorder \cite{Pal2024_ScientificReports, Panda2026,ref2,ref3,ref8,a13, Ghosh2025}.  One of the most remarkable consequences of disorder in magnetic systems is the emergence of the Griffiths phase (GP), originally proposed for diluted Ising ferromagnets \cite{Griffiths1969}. In this phase, short-range magnetically ordered clusters exist within a globally paramagnetic phase in the temperature range $T_C < T < T_G$, where $T_G$ is the Griffiths temperature. The presence of these clusters leads to non-analytic behavior of thermodynamic quantities and deviation from the Curie–Weiss law in magnetic susceptibility. In addition to Griffiths phase behavior, such magnetically clustered states can also give rise to a superparamagnetic (SPM) regime, particularly when the clusters become sufficiently small and thermally fluctuating, behaving like single-domain magnetic entities with no long-range order. Subsequent experimental and theoretical studies have established that the Griffiths phase is not limited to simple ferromagnets but is a universal feature of disordered magnetic systems \cite{CastroNeto1998,a3}. In manganites, the coexistence and competition between double-exchange (DE) and superexchange (SE) interactions, combined with intrinsic disorder, naturally lead to the formation of ferromagnetic clusters within the paramagnetic matrix above the magnetic transition temperature \cite{Dagotto2001}. This results in a Griffiths-like phase characterized by a downturn in inverse susceptibility and anomalous magnetic response. Moreover, depending on the size distribution and thermal stability of these clusters, a crossover from Griffiths phase to superparamagnetic like behavior may occur, where the clusters act as independent fluctuating magnetic moments.

Importantly, the Griffiths phase has also been observed in antiferromagnetic and frustrated systems, where disorder induces a distribution of competing exchange interactions. In such systems, even though the long-range ground state is antiferromagnetic, local regions can develop ferromagnetic correlations due to fluctuations in exchange interactions. For example, in the frustrated intermetallic compound GdFe$_{0.17}$Sn$_2$, structural defects and non-stoichiometry lead to random exchange interactions and the emergence of Griffiths phase behavior coexisting with antiferromagnetism \cite{Saha2025, Pal2019B, Phan2014, a4,a9}. Similarly, in U(Mn$_{1-x}$Cu$_x$)$_2$Ge$_2$, random substitution introduces structural disorder that generates short-range spin correlations, resulting in Griffiths-like behavior above the magnetic transition temperature. These observations clearly demonstrate that structural disorder is a key factor responsible for the emergence of Griffiths phase even in antiferromagnetic systems.

Further evidence of disorder-driven Griffiths phase comes from oxide systems such as double perovskites and doped manganites, where cation disorder modifies exchange pathways and promotes magnetic inhomogeneity \cite{Salamon2001}. The presence of competing magnetic interactions and structural disorder creates a complex energy landscape that favors the formation of magnetically correlated clusters above the ordering temperature. Therefore, the Griffiths phase in manganites can be understood as a manifestation of disorder-induced magnetic inhomogeneity arising from competing exchange interactions. 
Perovskite manganites with the general formula $R_{1-x}A_x$MnO$_3$ ($R$ = trivalent rare-earth ion, $A$ = divalent alkaline-earth ion) are prototypical strongly correlated electron systems in which charge, spin, orbital, and lattice degrees of freedom are strongly coupled, leading to a rich variety of electronic and magnetic phases \cite{Imada1998,Tokura2000, a5}. These materials exhibit complex phase diagrams including ferromagnetic metallic (FMM), charge-ordered antiferromagnetic (CO-AFM), G-type antiferromagnetic (G-AFM), and canted antiferromagnetic (CAFM) states, arising from the competition between different microscopic interactions. The fundamental mechanism governing the magnetic behavior of manganites is the competition between ferromagnetic DE, driven by the hopping of itinerant $e_g$ electrons between Mn$^{3+}$ and Mn$^{4+}$ ions \cite{AHMED2025114760,Zener1951,Anderson1955,a7}, and antiferromagnetic SE interactions between localized $t_{2g}$ spins via Mn–O–Mn bonds \cite{Goodenough1955,a6}. The delicate balance between these interactions makes manganites highly sensitive to carrier concentration, lattice distortions, and bandwidth, which can be effectively tuned by chemical substitution \cite{Dagotto2001,a8}. Replacing the A-site cation by a dopant leads to changes in several structural parameters such as the lattice constants ($a$, $b$, and $c$), unit cell volume, and in some cases, even a transition between different crystal structures. Hole doping introduces charge imbalance in the system, which is compensated by the conversion of some Mn$^{3+}$ ions into Mn$^{4+}$ ions, resulting in a mixed-valence state. The interaction between Mn$^{3+}$--O$^{2-}$--Mn$^{4+}$ ions gives rise to ferromagnetic (FM) double exchange (DE) interactions, leading to a metallic ground state. In contrast, Mn$^{3+}$--O$^{2-}$--Mn$^{3+}$ and Mn$^{4+}$--O$^{2-}$--Mn$^{4+}$ interactions favor antiferromagnetic (AFM) superexchange interactions, which result in an insulating ground state. Furthermore, structural factors such as lattice distortion, Mn$^{3+}$/Mn$^{4+}$ ratio, $\langle$Mn--O--Mn$\rangle$ bond angle, Mn--O bond length, octahedral rotation angle ($\theta$), and other related parameters significantly influence the physical properties of these systems. Doping-induced disorder can also lead to the emergence of novel magnetic phases, such as Griffiths singularity, superparamagnetic phase, which are typically absent in undoped compounds. In this context, the compound La$_{0.19}$Ca$_{0.81}$MnO$_3$ lies in a critical region of the phase diagram characterized by strong antiferromagnetic interactions and enhanced disorder effects. Despite extensive studies on manganites, the nature of Griffiths phase and its possible coexistence or crossover with superparamagnetic-like behavior in this highly doped regime remains relatively unexplored. In the present work, we investigate the existence and origin of the Griffiths phase in bulk La$_{0.19}$Ca$_{0.81}$MnO$_3$, with particular emphasis on the role of structural disorder, competing magnetic interactions, and short-range magnetic correlations in the paramagnetic region.

	
\section{Experimental Details}

{ Bulk polycrystalline $\mathrm{La}_{0.19}\mathrm{Ca}_{0.81}\mathrm{MnO}_{3}$ was prepared by a conventional high-temperature solid-state reaction route using a nominal weighed composition of $\mathrm{La}_{0.19}\mathrm{Ca}_{0.81}\mathrm{MnO}_{3}$.} High-purity La$_2$O$_3$ (99.99\%), CaCO$_3$ (99.99\%), and MnO$_2$ (99.9\%) were used as starting materials. Prior to weighing, La$_2$O$_3$ was preheated at 500~$^\circ$C for 12~h to remove moisture and absorbed carbonates. Stoichiometric amounts of the precursor powders were thoroughly mixed and ground in an agate mortar for several hours to ensure homogeneity. The mixed powder was calcined at 900~$^\circ$C for 24~h in air, reground, and then heated again at 1100~$^\circ$C for 24~h to promote complete solid-state reaction. The resulting powder was pressed into circular pellets and finally sintered at 1300~$^\circ$C for 36~h in air \cite{PAL2024120276}. After sintering, the pellets were furnace-cooled to room temperature.

Room-temperature phase analysis was carried out by powder X-ray diffraction (XRD) using a high-power Rigaku TTRAX-III diffractometer equipped with an $18\,\text{kW}$ rotating anode $\text{Cu-K}_\alpha$ radiation source ($\lambda = 1.5406\,\text{\AA}$) operating at $50\,\text{kV}$ and $300\,\text{mA}$. The instrument optical path utilized a cross-beam optics (CBO) unit providing a parallel beam geometry, a high-resolution graphite monochromator on the diffracted beam side, and a high-speed scintillation counter detector. { The Rietveld refinement was performed using FullProf in the $2\theta$ range $20^{\circ}$--$90^{\circ}$ with a step size of $0.02^{\circ}$ and a scan rate of $10^{\circ}$ min$^{-1}$. La and Ca were assigned to the common $4c$ Wyckoff site, and their site-occupancy factors were linearly constrained to full A-site occupancy, $\mathrm{SOF}_{\mathrm{La}}+\mathrm{SOF}_{\mathrm{Ca}} =\mathrm{Multiplicity}$. La and Ca shared identical fractional coordinates, and their isotropic thermal parameters were constrained such that $B_{\mathrm{La}}=B_{\mathrm{Ca}}$.}
This ultra-high-flux configuration provided exceptional counting statistics (peak intensities exceeding $10^5$ counts), which mathematically accounts for the low expected residual ($R_{\text{exp}} = 0.56\%$) obtained during the structural Rietveld refinement. Comparative profile checks performed at a reduced scan rate of $5^\circ\,\text{min}^{-1}$ yielded identical structural parameters, verifying that the $10^\circ\,\text{min}^{-1}$ configuration provides robust and optimized statistical accuracy free from truncation artifacts \cite{PAL2024102316}. The microstructure and local chemical composition of the sample were investigated using scanning electron microscopy coupled with energy-dispersive X-ray spectroscopy (SEM--EDS). High-magnification electron microscopy was employed to evaluate the surface morphology and microscale homogeneity. EDS spectra and elemental maps were acquired from a representative selected region using an accelerating voltage of 21~kV, a detector take-off angle of approximately $35.8^\circ$, and an energy resolution of 130.4~eV. The quantitative EDS data were analysed using standardless eZAF correction\cite{PAL2023170794} .
Magnetic measurements were performed using a Quantum Design SQUID-VSM magnetometer. The temperature dependence of magnetization, \(M(T)\), was measured in the range 5--400~K under magnetic fields up to 7~T using both zero-field-cooled (ZFC) and field-cooled (FC) protocols. For the ZFC measurement, the sample was cooled from 400~K to 5~K in zero field and the data were recorded during warming under the applied field, whereas in the FC mode the sample was cooled in the presence of the magnetic field. Isothermal magnetization, \(M(H)\), curves were recorded at selected temperatures across the magnetic transition region with the magnetic field varied between \(-7\) and \(+7\)~T to examine the nature of the magnetic transition and evaluate the magnetocaloric response.

	
\begin{figure*}
	\centering
	\includegraphics[width=.95\linewidth]{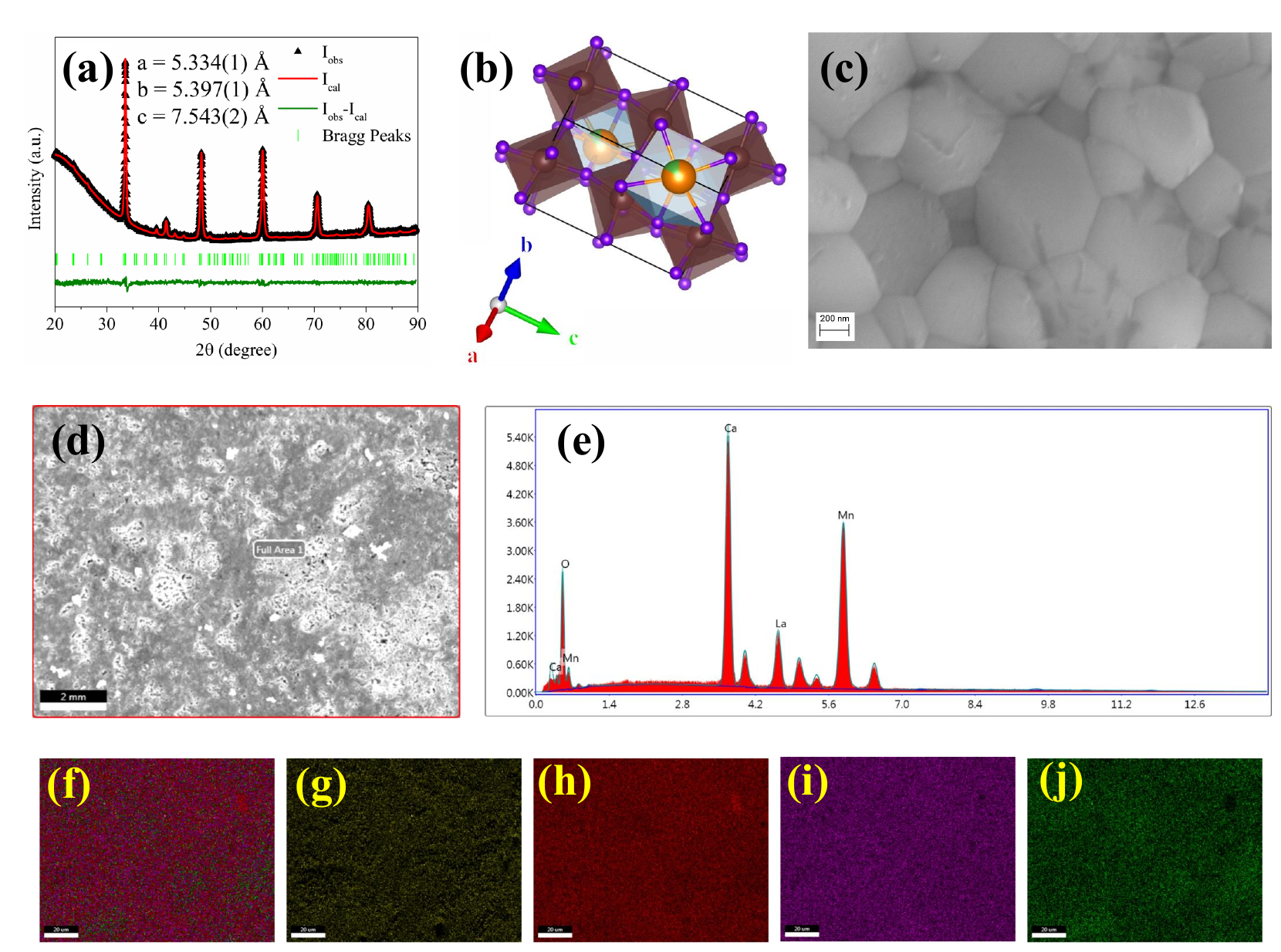}
	\caption{ (a) Room-temperature powder X-ray diffraction pattern of bulk La$_{0.19}$Ca$_{0.81}$MnO$_3$ together with the Rietveld refinement profile (red line), the difference curve (olive line), and the calculated Bragg reflection positions (vertical ticks). The refinement confirms a single-phase orthorhombic \textit{Pnma} structure. (b) Polyhedral representation of the perovskite crystal structure showing the La/Ca (green/orange) and MnO$_6$ (octahedra) coordination polyhedra. The orthorhombic distortion arises from cooperative tilting of the MnO$_6$ octahedra. (c) High-magnification electron micrograph. (d) Representative region selected for EDS measurement. (e) EDS spectrum collected from the selected area, confirming the presence of O, Mn, Ca, and La. (f)--(j) Elemental maps showing the spatial distributions of O, Mn, Ca, and La, respectively. The uniform elemental distributions indicate good chemical homogeneity and no detectable elemental segregation at the micrometre scale.}
	\label{fig:xrd}
\end{figure*}
	
\section{Results and discussion}

\subsection{Structural Analysis}

Figure~\ref{fig:xrd}(a) shows the room-temperature powder X-ray diffraction pattern of bulk La$_{0.19}$Ca$_{0.81}$MnO$_3$ together with the Rietveld refinement analysis~\cite{rietveld1969profile}. All reflections can be indexed using an orthorhombic structure with space group \textit{Pnma} (No.~62; Hermann–Mauguin symbol P2$_1$/n2$_1$/m2$_1$/a)~\cite{radaelli1997structural}, and no additional peaks arising from secondary phases are detected. The close agreement between the observed and calculated patterns, along with a small difference curve, confirms that the sample is single phase.
The Rietveld refinement was performed using the FullProf software~\cite{rodriguez1993recent} in the \(2\theta\) range \(20^\circ\)–\(90^\circ\) with a step size of \(0.02^\circ\) and a scan rate of \(10^\circ\)\,min\(^{-1}\). { The refinement yielded a profile-weighted residual $R_{\mathrm{wp}}=1.88\%$, an expected residual $R_{\mathrm{exp}}=0.56\%$, a goodness-of-fit $S=R_{\mathrm{wp}}/R_{\mathrm{exp}}=3.36$, and a phase reliability factor $R_{\mathrm{phase}}=1.74\%$. The small value of $R_{\mathrm{exp}}$ represents the statistical reference level of the profile refinement and is not interpreted as evidence of high sensitivity to the La/Ca occupancy.} 
The refined lattice parameters are \(a = 5.334(1)\)~\AA, \(b = 5.397(1)\)~\AA, and \(c = 7.543(2)\)~\AA, giving an orthorhombic unit-cell volume \(V = 217.1\)~\AA$^3$ and an X-ray density of 4.94~g\,cm$^{-3}$~\cite{radaelli1997structural,martin1999magnetic}.

It must be emphasized that because laboratory X-ray diffraction intensity is heavily dominated by electron-dense metal cations ($\text{La}$, $\text{Ca}$, $\text{Mn}$), the light oxygen atoms contribute marginally to the total structural scattering factor. Consequently, the individual oxygen fractional coordinates carry a higher level of statistical uncertainty compared to the cation parameters, which represents an intrinsic limitation of laboratory diffractometer systems compared to Neutron Powder Diffraction~\cite{radaelli1997structural}. To overcome this limitation and minimize unphysical correlations during the refinement loops, the oxygen positions were regularized within narrow limits based on standard orthorhombic parameters, and the anisotropic line shapes were smoothed using March-Dollase preferred orientation scaling~\cite{radaelli1997structural}. Furthermore, the Site Occupancy Factors ($\text{SOFs}$) for the mixed $\text{La}/\text{Ca}$ sublattice sharing the $4c$ site were kept as variable near the nominal synthetic values ($0.19/0.81$) to maintain stoichiometry constraints, and isotropic thermal displacement parameters ($B_{\text{iso}}$) were introduced to provide a complete structural representation as compiled in Table I. { The A site of the perovskite lattice was modelled using La$^{3+}$ and Ca$^{2+}$ on the common $4c$ Wyckoff position. The La and Ca occupancies were refined under the constraint of full A-site occupancy. Refinements initiated using different La/Ca ratios converged to essentially the same final fractions of approximately 0.19 and 0.81, indicating stability of the least-squares solution with respect to the tested starting values. These refined fractions are not interpreted as a four-decimal determination of the chemical composition.}
Mn$^{4+}$ ions fully occupy the 4$b$ site at \((0,0,\tfrac{1}{2})\), while oxygen atoms occupy one 4$c$ site (O1) and one 8$d$ site (O2). The fractional coordinates derived from the refinement are consistent with those expected for orthorhombic \textit{Pnma} manganites~\cite{radaelli1997structural,martin1999magnetic}.\par

\begin{table}[t]
\caption{Refined atomic coordinates, site occupancy factors (SOF), and isotropic thermal displacement parameters ($B_{\mathrm{iso}}$) for bulk polycrystalline La$_{0.19}$Ca$_{0.81}$MnO$_3$ at room temperature (Space Group: $Pnma$; $a = 5.334(1)$ \AA, $b = 5.397(1)$ \AA, $c = 7.543(2)$ \AA).}
\label{tab:structure}
\centering
\resizebox{\columnwidth}{!}{%
\begin{tabular}{ccccccc}
\hline
Atom & Site & $x$ & $y$ & $z$ & SOF & $B_{\mathrm{iso}}$ (\AA$^{2}$) \\
\hline
La/Ca & $4c$ & 0.00030(1) & 0.25000 & 0.03040(3) & $0.19/0.81^{a}$ & 0.65(3) \\
Mn    & $4b$ & 0.00000    & 0.00000 & 0.50000    & 1.0000              & 0.48(2) \\
O1    & $4c$ & 0.91400(2) & 0.25000 & 0.49200(1) & 1.0000              & 0.85(7) \\
O2    & $8d$ & 0.29700(3) & 0.97350(1) & 0.28300(4) & 1.0000            & 0.85(7) \\
\hline
\end{tabular}
}
\vspace{4pt}

\begin{minipage}{0.95\linewidth}
\small
\textit{Footnote:} { The La and Ca occupancies were refined under the constraint of full A-site occupancy. Refinements using different initial La/Ca ratios converged to essentially the same final values. The reported occupancies represent the stable solution of the constrained structural model and are not interpreted as a four-decimal chemical-composition determination.}
\end{minipage}

\end{table}

The local coordination around Mn and the corresponding Mn–O bond geometry were analyzed using the refined structure. Each Mn ion is octahedrally coordinated by six oxygen ions, forming slightly distorted MnO$_6$ octahedra. The inequivalent Mn–O bond lengths are 1.883~\AA\ (Mn–O1) and 1.941~\AA\ (Mn–O2), giving an average Mn–O bond length of \(\langle \mathrm{Mn–O} \rangle = 1.912\)~\AA. The corresponding Mn–O–Mn bond angles are 158.29° (Mn–O1–Mn) and 152.53° (Mn–O2–Mn), reflecting cooperative tilting of the MnO$_6$ octahedra within the \textit{Pnma} framework. The octahedral distortion parameter is \(\Delta = 2.02 \times 10^{-4}\), confirming that the MnO$_6$ units are only moderately distorted in this highly Ca-doped composition.

\begin{table}[h]
	\centering
	\caption{Selected bond lengths and bond angles for La$_{0.19}$Ca$_{0.81}$MnO$_3$ obtained from the refined structure.}
	\begin{tabular}{lc}
		\hline
		Bond parameter & Value \\
		\hline
		Mn–O1 (\AA) & 1.883 \\
		Mn–O2 (\AA) & 1.941 \\
		$\langle$Mn–O$\rangle$ (\AA) & 1.912 \\
		Mn–O1–Mn (deg) & 158.29 \\
		Mn–O2–Mn (deg) & 152.53 \\
		$\Delta$ & $2.02 \times 10^{-4}$ \\
		\hline
	\end{tabular}
	\label{tab:bond_parameters}
\end{table}

The orthorhombic distortion in La$_{0.19}$Ca$_{0.81}$MnO$_3$ can thus be ascribed to cooperative MnO$_6$ octahedral tilting combined with residual Jahn–Teller effects associated with the finite Mn$^{3+}$ content (\(\sim 18.75\%\)). Although the high Mn$^{4+}$ concentration in this strongly Ca-doped composition suppresses Jahn–Teller distortion relative to low-doped manganites, the remaining lattice distortion—with Mn–O–Mn angles \(\sim 155^\circ\) and modest octahedral tilting—is sufficient to support charge-order tendencies and antiferromagnetic interactions. The distortion in the structure is developed and it can be estimated by using the equations in the plane ($ab$ plane) and out-of-plane (along $c$ axis):
\begin{equation}
OS_{\parallel} = \frac{2(b - a)}{(b + a)}
\end{equation}
\begin{equation}
OS_{\perp} = \frac{2\left(a + b - \frac{c}{\sqrt{2}}\right)}{\left(a + b + \frac{c}{\sqrt{2}}\right)}
\end{equation}
for $a < \frac{c}{\sqrt{2}} < b$, where $a$, $b$, and $c$ are lattice parameters.
The calculated orthorhombic distortions, i.e., strain field parameters at $T = 300\,\mathrm{K}$, are $OS_{\parallel} \approx 0.012$ and $OS_{\perp} \approx 0.617$ respectively.
The orthorhombic distortion in La$_{0.19}$Ca$_{0.81}$MnO$_3$ expected to modify the magnetism of the studied compound \cite{Dhital2011}. { Because of peak overlap in the orthorhombic Pnma pattern and possible broadening from crystallite-size and microstrain effects, the independent information available for A-site occupancy refinement is limited. The different X-ray scattering powers of La and Ca, together with correlations among occupancy, scale, displacement, profile, and background parameters, further limit the precision of the refined La/Ca ratio. Refinements started from different La/Ca ratios converged to essentially the same final values, indicating stability with respect to the tested starting values; however, these values are not interpreted as a four-decimal determination of the chemical composition.}

To verify the chemical stoichiometry and cross-check the target composition, Energy Dispersive X-ray Spectroscopy ($\text{EDS}$) measurements were conducted on the synthesized pellets. The quantitative elemental data confirmed that the relative atomic ratios of $\text{La}$, $\text{Ca}$, and $\text{Mn}$ align tightly with the nominal 4-digit formulation of $\text{La}_{0.19}\text{Ca}_{0.81}\text{MnO}_3$ within standard experimental error margins. Figure~\ref{fig:xrd}(c) presents the high-magnification electron micrograph of the sample. The microstructure appears compact and spatially homogeneous within the investigated area, without obvious large secondary-phase particles or pronounced compositional contrast.
The region selected for EDS characterization is shown in
Fig.~\ref{fig:xrd}(d). The corresponding EDS spectrum in
Fig.~\ref{fig:xrd}(e) exhibits characteristic signals from La, Ca, Mn, and O, confirming the presence of all expected constituent elements in the La--Ca--Mn--O sample. Quantitative analysis of the full selected area gives atomic concentrations of 57.4\% O, 21.3\% Mn, 17.35\% Ca, and 3.95\% La. 
{   The EDS results are semi-quantitative and support an approximate composition of $\mathrm{La}_{0.19}\mathrm{Ca}_{0.81}\mathrm{MnO}_{3}$; they were not used to establish the La/Ca ratio at four-decimal precision.}
Figures~\ref{fig:xrd}(f)--\ref{fig:xrd}(j) show the elemental distributions of O, Mn, Ca, and La. All elements are distributed uniformly throughout the mapped region, with no detectable local enrichment or depletion of Mn, Ca, or La. The absence of obvious elemental segregation supports good chemical homogeneity at the micrometre scale and indicates that no secondary phase is detected within the spatial resolution and detection limit of EDS. {  The values 0.1875 and 0.8125 denote the nominal stoichiometric fractions used for weighing the starting materials, whereas the sample composition is reported as $\mathrm{La}_{0.19}\mathrm{Ca}_{0.81}\mathrm{MnO}_{3}$ in accordance with the semi-quantitative precision of EDS.}

\begin{figure*}
	\centering
	\includegraphics[width=.95\linewidth]{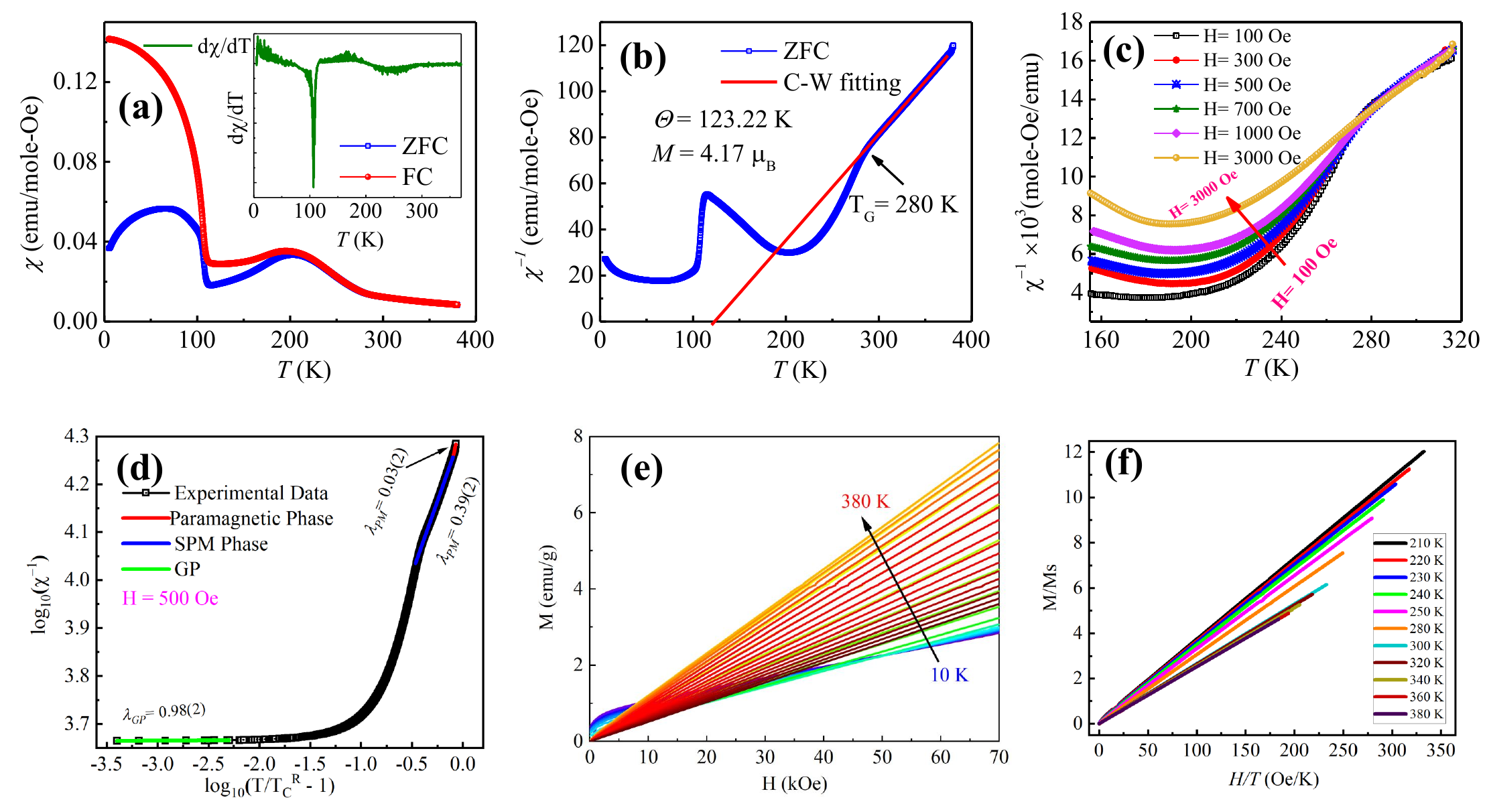}
	\caption{(a) Temperature dependence of magnetic susceptibility measured under zero-field-cooled (ZFC) and field-cooled (FC) conditions at an applied field of 500~Oe for La$_{0.19}$Ca$_{0.81}$MnO$_3$. The inset shows the temperature derivative \(d\chi/dT\), highlighting the magnetic anomalies. (b) Temperature-dependent inverse susceptibility ($1/\chi$) curve for this compound measured under an applied magnetic field of 500~Oe. The solid red line represents the Curie–Weiss (CW) fit. (c) $1/\chi$ under external magnetic fields of 100~Oe, 300~Oe, 500~Oe, 700~Oe, 1000~Oe, and 3000~Oe. The solid lines serve as guides to the eye, highlighting deviations from CW behavior. (d) Double-logarithmic plot of inverse susceptibility data against the reduced temperature, with solid red lines indicating linear fits in the Griffiths phase (GP) region. (e) Temperature evolution of the isothermal \(M(H)\) curves measured from 10 to 380~K for La$_{0.19}$Ca$_{0.81}$MnO$_3$ compound. (f) $M/M_{\mathrm{S}}$ (normalized magnetization) as a function of $H/T$ for the temperature range 210-380~K.}
	\label{mag}
\end{figure*}

\subsection{ Magnetization Study}

The temperature dependence of zero-field-cooled (ZFC) and field-cooled (FC) dc magnetization [$\chi(T)$] was measured in the temperature range of 5~K to 380~K under an applied magnetic field of $H = 500$~Oe, as shown in Fig.~\ref{mag}(a). A bifurcation between the ZFC and FC curves is observed below $T \sim 240$~K. Upon cooling, $\chi(T)$ shows a clear increase, indicating the gradual development of magnetic correlations.
Two distinct magnetic features are observed in the temperature dependence of the magnetic susceptibility. A broad anomaly appears around $T_1 \approx 210~\mathrm{K}$, which is attributed to an antiferromagnetic (AFM) transition. The relatively broad nature of this feature, along with the absence of a sharp singularity in $d\chi/dT$ (inset of Fig.~\ref{mag}(a)), suggests that it is associated with the gradual development of antiferromagnetic correlations rather than a well-defined second-order phase transition. A second, much sharper anomaly is observed at $T_2 \approx 100~\mathrm{K}$, which is clearly visible as a pronounced extremum in $d\chi/dT$. This feature marks the onset of long-range magnetic ordering. Considering the highly Ca-rich composition and its location near the CO-AFM/CAFM boundary, this transition is consistent with predominantly antiferromagnetic ordering, possibly accompanied by weak ferromagnetic canting or a small ferromagnetic component arising from magnetic phase coexistence. Below 240 K, a clear bifurcation between the FC and ZFC curves develops at 500~Oe, indicating significant thermomagnetic irreversibility. Such irreversibility points to competing magnetic interactions and may be associated with canted AFM order, ferromagnetic clusters, or weak glassy spin freezing at low temperature.
At higher field, the temperature-dependent magnetization measured under ZFC and FC conditions at \(H = 70\)~kOe [see in supplementary Figure S1] shows that the bifurcation is strongly suppressed and the anomaly near \(T_2\) becomes broader. This field-induced reduction of irreversibility indicates that the low-temperature frustrated magnetic state is highly field sensitive. The applied field tends to stabilize the ferromagnetic component, weaken the effects of magnetic competition, and suppress domain-related or glassy behavior.

\par The minimum in the $d\chi/dT$ versus $T$ plot extends over a broad temperature range of approximately 130~K (from 190~K to 300~K). This indicates that the transition in that region is not of a long-range ferromagnetic nature and suggests a possible presence of short-range magnetic interactions above the Néel temperature ($T_N$). Therefore, to better understand the magnetic behavior in the high-temperature region approaching the phase transition, as well as the nature of the long-range ordering and its universality class, a critical analysis around the antiferromagnetic phase transition temperature of the compound has been carried out for this compound.
The inverse susceptibility as a function of temperature was plotted to understand the magnetic nature of the compound La$_{0.19}$Ca$_{0.81}$MnO$_3$. Figure~\ref{mag}(b) illustrates the temperature dependence of the inverse susceptibility. A downward deviation is observed below 280~K, while above this temperature, $\chi^{-1}(T)$ follows typical Curie-Weiss behavior. The Curie-Weiss fitting yields a Weiss temperature $\theta_W = 123.22$~K, indicating a dominant ferromagnetic (FM) interaction.
Figure~\ref{mag}(c) shows the temperature dependence of the inverse susceptibility under different applied magnetic fields ranging from 100~Oe to 3~kOe. A significant downward deviation is observed at temperatures well above the magnetic ordering temperature. This deviation suggests the formation of short-range magnetic clusters within the paramagnetic matrix~\cite{Ghosh2025,PAL2024140211,a11, Dutta2025,PAL2025112814,Pal2026_JPCC,a10}.
Furthermore, the $\chi^{-1}(T)$ plot exhibits a deviation from linearity below approximately 280~K. This behavior is characteristic of the Griffiths phase (GP), as described by Griffiths and Bray’s theory, where the system deviates from ideal paramagnetic behavior without establishing long-range magnetic order~\cite{Griffiths1969}. Therefore, a detailed investigation of the magnetic properties of La$_{0.19}$Ca$_{0.81}$MnO$_3$ has been carried out to better understand its magnetic behavior. In a different way, the deviation in $\chi^{-1}(T)$ from the Curie--Weiss (CW) line is very large, and the temperature range of the Griffiths phase (GP), normalized with respect to $T_N$, is calculated as:
\begin{equation}
\mathrm{GP}(\%\,T_N) = \left[ \frac{T_G - T_N}{T_N} \right] \times 100
\end{equation}
which is also high, i.e., $\sim 33\%$~\cite{Saha2025}. 

To confirm the presence of the Griffiths phase below 280~K, a detailed GP analysis was performed. The Griffiths singularity is characterized by the exponent $\lambda$, determined from the modified Curie–Weiss relation,
\[
\chi^{-1}(T) \propto (T - T_C^R)^{(1 - \lambda)},
\]
where $0 \leq \lambda < 1$, and $T_C^R$ represents the critical temperature at which $\chi$ diverges.
The exponent $\lambda$ quantifies the deviation from ideal Curie–Weiss behavior, with $\lambda \to 0$ in the paramagnetic (PM) regime and $\lambda > 0$ in the Griffiths phase (GP) region \cite{Ghosh2025, nasir2020}. The reference temperature $T_C^R$ was determined using a self-consistent fitting procedure. Initially, the inverse susceptibility data in the high-temperature paramagnetic region were analyzed by assuming $T_C^R = T_N$. The value of $T_C^R$ was then systematically adjusted, and the physically meaningful value was identified as that for which the exponent obtained in the paramagnetic regime approaches zero, indicating recovery of ideal Curie–Weiss behavior. This approach is justified by the fact that above the Griffiths temperature $T_G$, the system transitions into a conventional paramagnetic state that obeys the Curie–Weiss law. The final value of $T_C^R$ obtained through this procedure was subsequently used to extract the Griffiths exponent $\lambda_{\rm GP}$ \cite{ref3}.
Figure~\ref{mag}(d) presents the log–log plot of $\log_{10}(\chi^{-1})$ versus $\log_{10}(T/T_C^R - 1)$, along with the corresponding linear fit. The slope yields the exponent $\lambda_{GP} \approx 0.98(2)$, which reflects a strong Griffiths singularity in this compound~\cite{Karmakar2013, nasir2020}. A large $\lambda_{GP}$ value indicates pronounced FM clustering and significant magnetic inhomogeneity. Interestingly above the Griffiths temperature $T_G$, the exponent $\lambda_{PM}$ is not equal to zero but $\lambda_{PM} \approx 0.39(2)$ , and above 370 K temperature, the exponent $\lambda_{PM} \approx 0.0(2)$.

To examine the possible reason why $\lambda_{PM}$ does not become zero for $T > T_G$, the normalized magnetization ($M/M_s$) was plotted as a function of $H/T$ over the temperature range 210-380~K, as shown in figure~\ref{mag}(f) using the isothermal magnetization curves is presented in figure~\ref{mag}(e). In the case of superparamagnetic behavior, these curves are expected to collapse onto a single universal curve~\cite{Roy2008}. However, a deviation from scaling is observed in the temperature range 210--280~K, whereas the curves collapse reasonably well in the range 300-370~K. This behavior provides evidence for the presence of a superparamagnetic-like phase at temperatures above $T_G$, persisting up to approximately 370~K.

To further verify the presence of the GP and its correlation with magnetic disorder, the $\chi(T)$ data under different magnetic fields (100–10000~Oe) were analyzed using the theoretical model proposed by Galitski \textit{et al.}~\cite{Galitski2004, Dutta2025}. According to this model, the magnetization in the GP regime follows the relation
\begin{equation}
	M(T, H) \propto \exp[-C'(T/H)],
\end{equation}
where $C' = k_B / \mu_{\text{ave}}$, with $k_B$ being the Boltzmann constant and $\mu_{\text{ave}}$ representing the average magnetic moment of FM clusters. The fits, shown in figure~\ref{fig:MCE}(a), accurately capture the experimental $\chi(T)$ data within the GP region. The figure~\ref{mag}(b) demonstrates that the fitted $C'$ values increase with the applied magnetic field, implying a reduction in $\mu_{\text{ave}}$. This field dependence corroborates the presence of a robust Griffiths phase in La$_{0.19}$Ca$_{0.81}$MnO$_3$ compound, originating from competing AFM and FM interactions and inherent structural disorder. The emergence of GP behavior can thus be regarded as a natural manifestation of the same underlying mechanisms that give rise to the observed spin-glass-like magnetic state~\cite{Dutta2025, Ghosh2025}.
Arrott--Belov plots ($M^2$ versus $H/M$) show strong nonlinearity and no positive $M^2$-axis intercepts (Supplementary Figure S2), ruling out conventional long-range ferromagnetic order.

It is instructive to contrast this behavior with the low Calcium-doping region of the $\text{La}_{1-x}\text{Ca}_x\text{MnO}_3$ phase diagram. For instance, Jiang et al. \cite{Jiang2009} demonstrated a sharp onset of concentration-dependent Griffiths phase behavior, where GP-like features are completely absent at $x = 0.18$ but abruptly stabilize at $x = 0.19$ due to local variations in exchange couplings near the metal-insulator boundary. On the opposite end of the spectrum, in our highly Ca-doped $\text{La}_{0.19}\text{Ca}_{0.81}\text{MnO}_3$ compound, a similar sensitivity to local environment emerges. Here, the intense quenched disorder on the $A$-site disrupts the nominal long-range antiferromagnetic superexchange paths, giving rise to local ferromagnetic entities within the paramagnetic matrix well above $T_{\text{N}}$. This underscores exchange randomness as a universal driver for cluster states at both extremes of the manganite compositional matrix.\\

\subsection{Magnetocaloric Effect}
\begin{figure*}[htbp]
	\centering
	\includegraphics[width=0.95\linewidth]{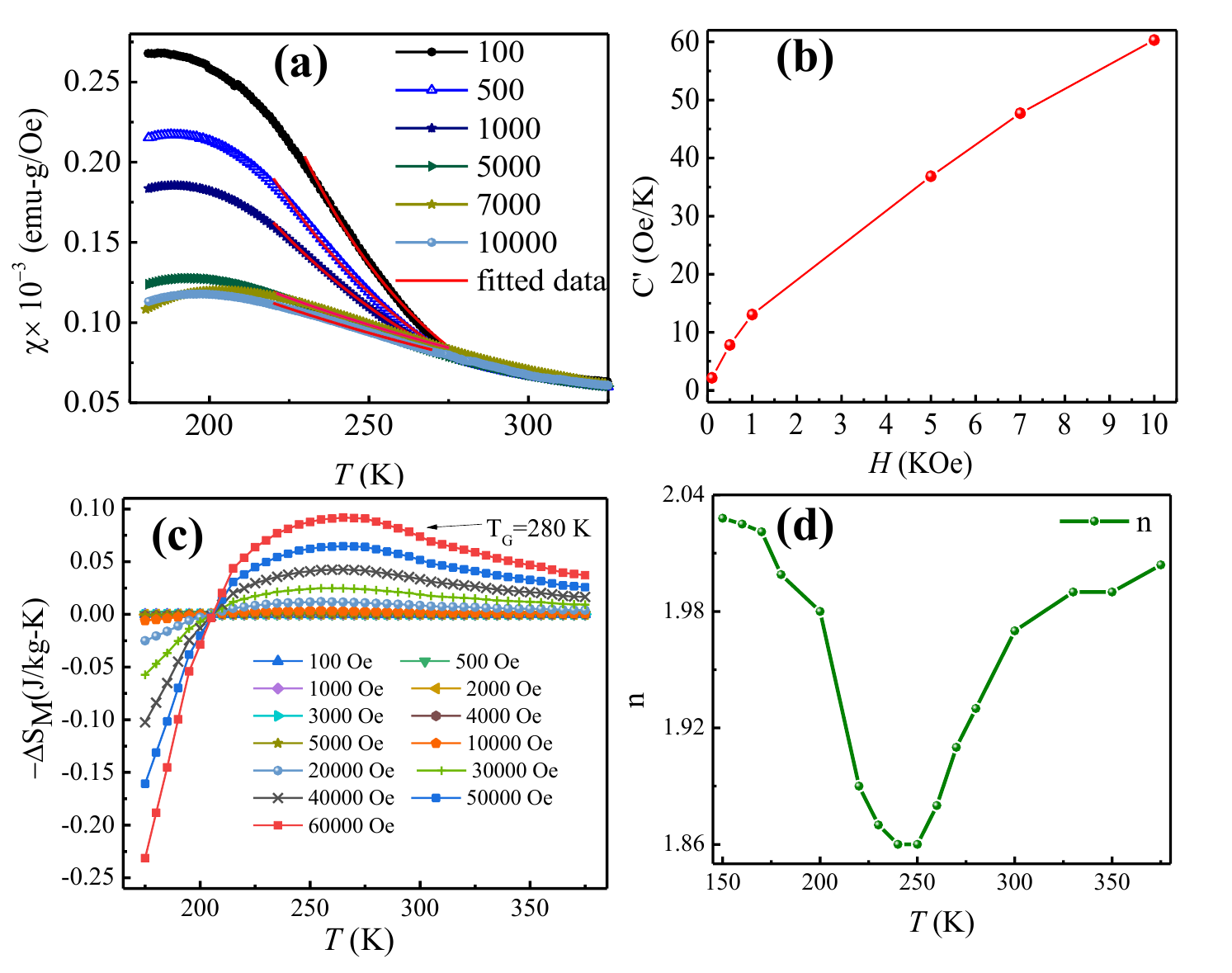}
	\caption{(a) Galitski model fitting for magnetic fields ranging from 100~Oe to 10000~Oe. (b)The variation of the estimated constant $C'$ with the applied magnetic field.  (c) Temperature dependence of the magnetic entropy change ($-\Delta S_M$) under different applied magnetic fields ranging from 100 Oe to 60 kOe. The entropy change exhibits a broad peak around the magnetic transition temperature $T_G \approx 280$ K. (d) Temperature dependence of the field exponent $n$, obtained from the relation $\Delta S_M \propto H^n$ for La$_{0.19}$Ca$_{0.81}$MnO$_3$ compound.}
	\label{fig:MCE}
\end{figure*}

Figure~\ref{fig:MCE}(c) shows the temperature dependence of the magnetic entropy change, $-\Delta S_{\mathrm{M}}$, for La$_{0.19}$Ca$_{0.81}$MnO$_3$ under different applied magnetic fields. The magnetic entropy change was calculated from isothermal $M(H)$ data shown in Fig. \ref{mag}(e) using the Maxwell relation,
\begin{equation}
	\Delta S_{\mathrm{M}}(T,H) = -\int_{0}^{H} \left( \frac{\partial M}{\partial T} \right)_{H} dH,
\end{equation}
with numerical evaluation via the standard finite-difference approximation between successive isotherms\cite{Ghosh2023,10.1039/d4tc00803k,
Pal_2024, a12}. The $-\Delta S_M(T)$ curves exhibit a systematic increase in magnitude with increasing magnetic field, confirming the intrinsic magnetic origin of the entropy change.

At lower temperatures ($T < 200$ K), $-\Delta S_M$ shows negative values, indicating the presence of an inverse magnetocaloric effect and aslo provides the evidence of AFM transition at this temperature for this compounds. With increasing temperature, $-\Delta S_M$ changes sign and develops a broad positive peak centered around 280 K. The peak magnitude increases progressively with field, reaching its maximum under 60 kOe. The broad nature of the entropy change curve suggests a distributed magnetic transition rather than a sharp first-order transition. The maximum entropy change occurs in the vicinity of the magnetic transition temperature, identified as $T_G \approx 280$ K. Close to $T_G$,$-\Delta S_M$ exhibits an well-defined peak, often seen in materials with ferromagnetic type ordering. The continuous evolution of $-\Delta S_M$ with temperature and field, along with the absence of any abrupt discontinuity, indicates that the magnetic transition is of second-order character. Furthermore, the broad operating temperature window of $-\Delta S_M$ highlights the potential applicability of the material for magnetic refrigeration over an extended temperature range.
To gain further insight into the magnetic interactions, the field dependence of the magnetic entropy change was analyzed using the scaling relation
\begin{equation}
\Delta S_M \propto H^n,
\end{equation}
where $n$ is the field exponent. For the paramagnetic region and antiferromagnetically ordered states, the expected value of $n$ is 2 \cite{Ghosh2024}. In the superparamagnetic state, the value of $n$ should be close to 2, whereas in the Griffiths phase it is expected to be below 2. 
The temperature dependence of $n$ is shown in figure~\ref{fig:MCE}(d). In the Neel-temperature region ($T_N\approx 210$), the exponent $n$ approaches a value close to 2, which is consistent with conventional antiferomagnetic behavior. As the temperature increases toward the Griffith's transition region, $n$ exhibits a pronounced deviation from this value and decreases, reaching a minimum of $\sim 1.86$ around 250 K. Such a reduction in $n$ above $T_G$ is indicative of the presence of short-range ferromagnetic correlations within the paramagnetic matrix. This anomalous behavior is characteristic of a Griffiths-like phase, where spatially distributed ferromagnetic clusters persist above the long-range ordering temperature due to intrinsic disorder, leading to non-mean-field behavior.

At higher temperatures ($T \gtrsim 300$ K), the exponent $n$ gradually recovers and approaches values very close to 2, with reduced temperature dependence. This behavior suggests the onset of a superparamagnetic-like regime, where thermally fluctuating magnetic clusters behave as non-interacting entities, resulting in a restoration of mean-field-like scaling. The collapse of $n$ toward $\sim 2$ in this temperature range is consistent with superparamagnetic dynamics of finite-sized ferromagnetic clusters.
The asymmetric variation of $n$ across the transition temperature, together with its suppressed values above $T_G$ and recovery toward $n \approx 2$ at higher temperatures, provides strong evidence for the coexistence of Griffiths-like behavior and a superparamagnetic-like phase in the system.

In summary, the compound La$_{0.19}$Ca$_{0.81}$MnO$_3$ exhibits a complex temperature-dependent magnetic evolution governed by competing ferromagnetic and antiferromagnetic interactions, along with intrinsic structural disorder.  
At high temperatures ($T > 370$~K), the system behaves as a conventional paramagnet (PM), where magnetic moments are randomly oriented and follow Curie-Weiss behavior. Upon cooling below $\sim 370$~K, short-range magnetic correlations begin to develop, giving rise to a superparamagnetic-like (SPM) regime characterized by thermally fluctuating magnetic clusters embedded within a paramagnetic matrix. Further cooling below the Griffiths temperature ($T_G \approx 280$~K) leads to the emergence of a Griffiths phase (GP), where strong magnetic inhomogeneity develops due to the formation of correlated FM clusters within the PM background. This is supported by the deviation of inverse susceptibility from Curie--Weiss behavior and a large Griffiths exponent, Galitski model fitting and power law analysis indicating pronounced disorder and cluster formation. Near the Néel temperature ($T_N \approx 210$~K), the system undergoes a broad antiferromagnetic transition, reflecting the gradual development of AFM correlations rather than a sharp phase transition.

\section{Conclusion}

In conclusion, we have demonstrated the robust emergence of a Griffiths phase and superparamagnetic-like behavior in bulk La$_{0.19}$Ca$_{0.81}$MnO$_3$, a system situated in the La-doped antiferromagnetic regime. A clear deviation from Curie-Weiss behavior, together with a field-dependent downturn in inverse susceptibility below $T_G \approx 280$ K, provides direct evidence for the formation of a Griffiths phase characterized by short-range ferromagnetic clusters embedded within a paramagnetic matrix. The extracted Griffiths exponent $\lambda_{\mathrm{GP}} \approx 0.98$ reveals a strong Griffiths singularity, reflecting pronounced magnetic inhomogeneity and the breakdown of conventional mean-field behavior. At high temperatures, the system exhibits conventional paramagnetic behavior, which on cooling evolves into a superparamagnetic-like regime characterized by thermally fluctuating magnetic clusters. Importantly, the presence of Griffiths phase in a predominantly antiferromagnetic system demonstrates that disorder-driven cluster formation is not restricted to ferromagnetic regimes but represents a universal feature of correlated oxides with competing interactions. The magnetocaloric and scaling analyses further support this picture, revealing nontrivial field dependence and signatures of short-range correlations across the Griffiths regime and above  $T_G \approx 280$ K upto 370K, superparamagnetic-like behaviour. Overall, this work establish A-site valence engineering as a powerful tool for tuning Griffiths phase and superparamagnetic-like behavior and open new pathways for controlling emergent magnetic functionality in complex oxides.

	\begin{acknowledgments}
The authors gratefully acknowledge Rui Wu and the South China University of Technology, China, for the fellowship and other assistance. They also acknowledge  the Saha Institute of Nuclear Physics, DAE, India, for providing support. The authors also thanks Afsar Ahmed for his assistance in different experimental understandings.
		
	\end{acknowledgments}

	\section*{Data Availability Statement}

The data that support the findings of this study are available from the corresponding author upon reasonable request.
	
	\bibliographystyle{apsrev4-2}
\bibliography{ref}
	
\end{document}